\documentclass[a4paper,11pt]{article}
\pdfoutput=1 

\usepackage{jcappub} 

\usepackage[T1]{fontenc} 
\usepackage{orcidlink}
\DeclareRobustCommand{\VAN}[3]{#2}
\let\VANthebibliography\thebibliography
\def\thebibliography{\DeclareRobustCommand{\VAN}[3]{##3}\VANthebibliography}
\usepackage{graphicx}	
\usepackage{amsmath}	
\usepackage{subcaption}

\title{\boldmath Analyzing Line-of-sight selection biases in galaxy-scale strong lensing with external convergence and shear}

\author[a,b,1]{Xianzhe TZ Tang\orcidlink{0009-0007-3185-7030},\note{Corresponding author.}}
\author[a]{Simon Birrer\orcidlink{0000-0003-3195-5507},}
\author[c,d,2]{Anowar J. Shajib\orcidlink{0000-0002-5558-888X},\note{NHFP Einstein Fellow}}
\author[a]{Narayan Khadka\orcidlink{https://orcid.org/0000-0001-5512-2716},}
\author[]{the LSST Strong Gravitational Lensing Science Collaboration,}
\author[]{and the LSST Dark Energy Science Collaboration}

\affiliation[a]{Department of Physics and Astronomy, Stony Brook University, Stony Brook, NY 11794, USA.}
\affiliation[b]{Boston University Department of Astronomy, 725 Commonwealth Ave, MA 02215, Boston USA}
\affiliation[c]{Kavli Institute for Cosmological Physics, University of Chicago, Chicago, IL 60637, USA}
\affiliation[d]{Center for Astronomy, Space Science and Astrophysics, Independent University, Bangladesh, Dhaka 1229, Bangladesh}

\emailAdd{tztang@bu.edu}
\emailAdd{simon.birrer@stonybrook.edu}

\abstract{The upcoming Vera Rubin Observatory Legacy Survey of Space and Time (LSST) will dramatically increase the number of strong gravitational lensing systems, requiring precise modeling of line-of-sight (LOS) effects to mitigate biases in lensing observations and cosmological inferences. We develop a method to construct joint distributions of external convergence ($\kappa_{\text{ext}}$) and shear ($\gamma_{\text{ext}}$) for strong lensing LOS by aggregating large-scale structure simulations with high-resolution halo renderings and non-linear correction. Our approach captures both smooth background matter and perturbations from halos, enabling accurate modeling of LOS effects. We apply non-linear LOS corrections to $\kappa_{\text{ext}}$ and $\gamma_{\text{ext}}$ that address the non-additive lensing effects caused by objects along the LOS in strong lensing. We find that, with a minimum image separation of $1.0^{\prime \prime}$, non-linear LOS correction due to the presence of a dominant deflector slightly increases the ratio of quadruple to double lenses; this non-linear LOS correction also introduces systematic biases of $\sim 0.1\%$ for galaxy–AGN lens in the inferred Hubble constant ($H_0$) if not accounted for. We also observe a $0.66\%$ bias for galaxy–galaxy lenses on $H_0$, and even larger biases $1.02\%$ for galaxy–AGN systems if LOS effects are not accounted for. These results highlight the importance of LOS for precision cosmology. The publicly available code and datasets provide tools for incorporating LOS effects in future analyses.}


\begin{document}
\maketitle
\flushbottom

\section{Introduction}

Strong gravitational lensing, where a massive object bends light rays to produce multiple images of a background source, is a powerful tool in astrophysics and cosmology. For example, strong lensing has been used to probe the nature of dark matter \citep[e.g.,][]{gilman2019probingdarkmatter}, to measure the Hubble constant \(H_0\) through time delays of multiply-imaged time-variable sources \citep[e.g.,][]{birrer2020tdcosmo, shajib2023tdcosmo}, to explore the internal structure of galaxies \citep[e.g.,][]{shajib2021massive}, and to constrain cosmological models using large samples of galaxy-galaxy lensing events \citep[e.g.,][]{chen2019assessing, li2024cosmology, Shajib:2024}.

The efficiency of creating strong lensing phenomena depends not only on the properties of the primary lensing galaxy, but also on the mass distribution along the line of sight (LOS). Mass over-densities, such as intervening galaxies or clusters, enhance the lensing convergence, increasing the likelihood of multiple images or highly magnified sources. Conversely, matter under-densities reduce this probability. Hence, LOS structures introduce selection effects that influence the observed lensing events. A further complication is that the first-order LOS convergence is not observable due to the mass-sheet degeneracy \citep[MSD;][]{falco1985model}. Hence, not well-characterized LOS of strong lenses can bias statistical analyses.

Understanding how LOS structures affect lensing observations is crucial, especially in the context of precision cosmology with upcoming large-scale surveys like the Vera C. Rubin Observatory's Legacy Survey of Space and Time (LSST), \emph{Euclid}, and the Nancy Grace Roman Space Telescope. Accurate modeling of LOS effects is necessary to account for biases in the inferred properties of lenses and cosmological parameters.

Early studies often assumed a smooth universe, neglecting inhomogeneities along the LOS \citep{ehlers2005gravitational}. To account for LOS effects, the tidal approximation introduces external convergence (\(\kappa_{\text{ext}}\)) and shear (\(\gamma_{\text{ext}}\)) to the lensing potential \citep{schneider1997tidal}. These LOS effects can significantly influence lensing observables, such as the ratio of double to quadruple images \citep{keeton1997shear}, with the alignment between the external shear and the lens galaxy ellipticity playing a substantial role \citep{oguri2010gravitationally}. External convergence effects can magnify or de-magnify the apparent size of lensing features, impacting detection thresholds and potentially introducing biases due to the mass-sheet degeneracy \citep[MSD;][]{falco1985model}. Moreover, LOS effects directly influence measurements of \(H_0\) in time-delay cosmography \citep{suyu2010dissecting, mccully2017quantifying, li2021impact} as external convergence transforms as a MSD and hence is a non-observable in strong lensing.


Various approaches have been explored to model LOS effects. Theoretical methods include Monte Carlo simulations assigning probability distributions for external shear and convergence \citep{oguri2010gravitationally}, and ray-tracing through cosmological simulations \citep{suyu2010dissecting, collett2016observational}. Empirical methods involve weighted galaxy number counts along the LOS to estimate external convergence \citep{suyu2010dissecting, fassnacht2011galaxy, wells2023tdcosmo}, and machine learning techniques using photometric data \citep{park2023hierarchical}. These methods require a prior distribution of LOS structure based on analytical or numerical simulations. However, these presented methods may not fully capture the non-linear interactions and contributions from individual halos along the LOS, especially those near the dominant lens.

In strong lensing, the presence of structures along the LOS can introduce non-linear effects into the lensing potential. General non-linearities in the vicinity of the main lens can arise from complex mass distributions near critical lines or from multiple structures interacting with the same light ray \citep{fleury2021line}. While such effects have been shown to bias parameter recovery and may act as nuisance parameters \citep{Duboscq_2024}, current observations typically lack the precision required to robustly measure them. The focus of our work is the non-linear coupling between otherwise linear LOS perturbations (e.g., shear and convergence) and the main deflector. The presence of a strong lens bends the light rays, and linearity and the Born approximation can only be assumed between observer and deflector, and deflector and source, but not along the full path between observer and source. Including the the presence of the dominant deflector in the LOS lensing  
leads to a first-order correction in the lensing efficiency, effectively rescaling the shear and convergence contributions \citep{fleury2021line}. Ignoring this non-linearity can introduce systematic uncertainties, including biases in inferred lens model parameters and the Hubble constant \citep{Johnson_2024}. By addressing this effect, we refine the treatment of LOS perturbations to improve the accuracy of strong lensing analyses.

An accurate prediction of the LOS effect in strong lensing is different from predicting large-scale structure shear and convergence maps used for most other applications: (i) The non-linear coupling of the main deflector requires the calculation of linear distortions conditional on both source and deflector redshift. (ii) Since strong lensing features occur on scales of $\sim$ $1''$, the prediction of the linear distortions have to match this resolution, effectively requiring very high-resolution maps, both in ray-tracing and in the predicted matter fluctuations.

In this work, we develop a method that constructs realistic joint distributions of external convergence and external shear (\(\kappa_{\text{ext}}\), \(\gamma_{\text{ext}}\)) for specific deflector and source redshifts. By aggregating large-scale structure simulations with high-resolution halo renderings and applying the non-linear correction coupling the deflector with the LOS, we capture both the smooth background matter distribution and the perturbations from individual halos along the LOS. This leads to more accurate modeling of LOS effects on lensing observables, improving the precision of lensing efficiency estimations and reducing potential biases in cosmological parameter estimation.

Furthermore, we provide an open-source tool that allows for the adoption of different cosmological models and fast sampling of LOS quantities across various redshifts. This flexibility facilitates large-scale analyses anticipated with upcoming surveys by efficiently accounting for LOS effects without the need for detailed individual analyses of each lensing system. Our methodology extends previous theoretical predictions by incorporating non-linear LOS effects and offers a practical solution for enhancing the accuracy of lensing studies and extracting cosmological information from extensive datasets. The tool is integrated into the LSST strong lensing simulation pipeline (\textsc{SLSim}, \cite{slsim}).

\section{Methodology}
A key requirement to assess strong lensing selections are accurate predictions on the external convergence (\( \kappa_{\text{ext}} \)) and shear (\( \gamma_{\text{ext}} \)) distributions at the scale of a strong lensing system and its dependence on deflector and source redshift.


Large-scale cosmological simulations of the matter density field are not able to predict accurate (\( \kappa_{\text{ext}} \) and \( \gamma_{\text{ext}} \) distributions at the scale of $1^{\prime\prime}$, a typical Einstein radius of a strong lensing system.
Instead of directly using numerical simulations with much increased resolution at a large cosmological volume, we have chosen a hybrid approach which makes use of cosmological simulations at large scales (see Section~\ref{LSS_method}), in tandem with a halo model approach on small scales (see Section~\ref{halos_method}).
The larger scales are derived from a large-scale structure simulation, which computes a low-resolution, yet realistic map of \( \kappa_{\text{ext}} \) and \( \gamma_{\text{ext}} \) by simulating the cosmological evolution of the matter power spectrum. This approach ensures a smooth and statistically correct large-scale representation. The second component involves high-resolution halo renderings, where we use ray-tracing techniques to compute \( \kappa_{\text{ext}} \) and \( \gamma_{\text{ext}} \) for individual halos. Moreover, to account for the fact that objects along the line of sight contribute non-linearly to the overall lensing effect based on their positions relative to the observer, deflector, and source, we incorporate non-linear line-of-sight corrections (see Section~\ref{los_nonlinear_correction_method}). By aggregating these two components, we capture both over-dense and under-dense regions (see Section~\ref{sec:method_Integration}). After generating the LOS distribution, we perform Monte Carlo simulations to obtain the distributions of \( \kappa_{\text{ext}} \) and \( \gamma_{\text{ext}} \) for specific deflector redshifts (\( z_\text{d} \)) and source redshifts (\( z_\text{s} \)). These distributions are then integrated into our simulation pipeline, \textsc{SLSim} \citep{slsim}, a comprehensive tool designed to generate realistic populations of strongly lensed systems under observational conditions similar to upcoming surveys like the LSST, allowing us to examine the selection effects of external convergence and shear thoroughly (see Section~\ref{sec:bias_evaluating}).

\subsection{Constructing realistic \( \kappa_{\text{ext}} \) and \( \gamma_{\text{ext}} \)\label{sec:building the kg}}

To accurately construct realistic \( \kappa_{\text{ext}} \) and \( \gamma_{\text{ext}} \) maps of strong lensing, we adopt an approach that combines insights from large-scale structure simulations and high-resolution halo rendering. Large-scale structure simulations (Section~\ref{LSS_method}) provide full-sky, low-resolution maps of convergence and shear, capturing the overall statistical properties. The halo rendering simulation (Section~\ref{halos_method}) delivers high-resolution maps within the light cone.  Furthermore, the non-linear LOS Correction (Section~\ref{los_nonlinear_correction_method}) accounts for the fact that, in strong lensing cases with a dominant lens, objects at different positions relative to the main deflector contribute differently to the total lensing signal.

\begin{figure}
    \centering
    \includegraphics[width=1\linewidth]{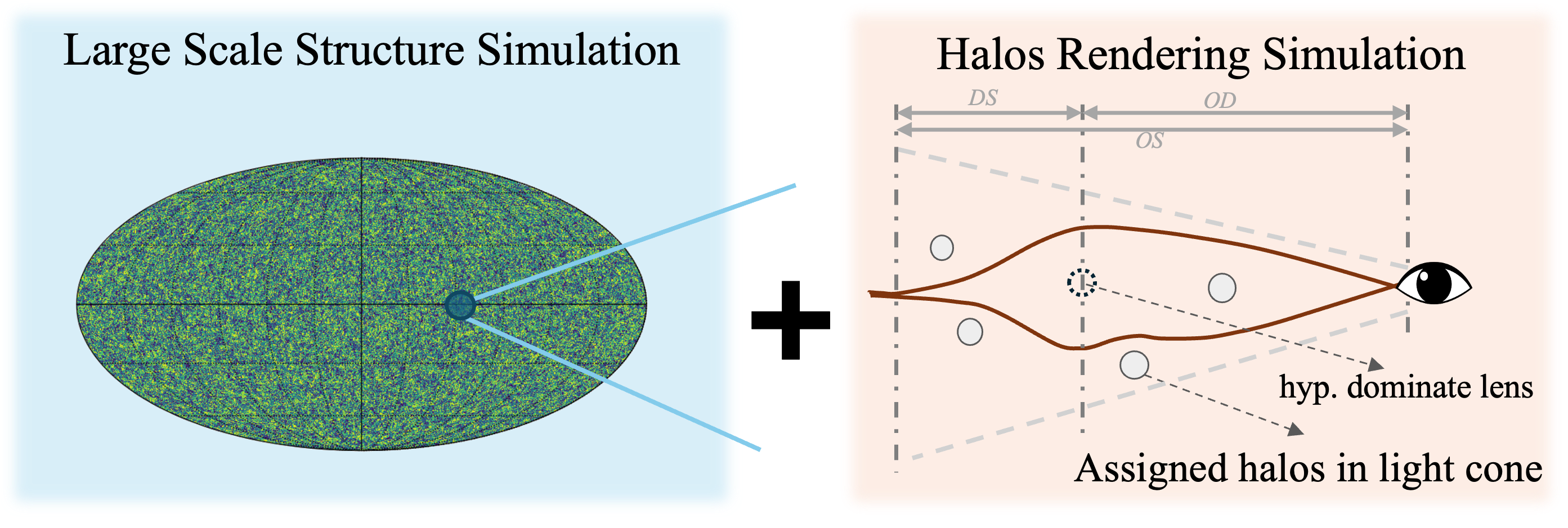}
    \caption{ For calculating a joint external convergence and shear maps, we aggregate large-scale structure simulation using \textsc{GLASS} package \citep{GLASS}, which computes a low-resolution, yet realistic $\kappa_{\text{LSS}}$ and $\gamma_{\text{LSS}}$ by simulating matter density shell by shell. This approach ensures a smooth background representation. On top of it high-resolution halos renderings, using ray-tracing techniques to compute $\kappa_{\text{halos}}$ and $\gamma_{\text{halos}}$ as sub-structure.}
    \label{fig:combine_lss_los}
\end{figure}

\subsubsection{Large-scale structure simulation \label{LSS_method}}
In this paper, we employ a large-scale structure simulation based on the matter power spectrum to simulate low-resolution lensing maps. Using the \textsc{GLASS} package \citep{GLASS}, we construct a full-sky light cone filled with discretized matter field, arranged in a series of nested shells. Weak gravitational lensing quantities, including convergence and reduced shear, are then computed as integrals over all lower-redshift matter shells.\citep{GLASS} This method provides a representation of the universe's large-scale structure. Note the full-sky simulation is distinct from the later halo rendering method (Section~\ref{sec:halo_rendering}), which focuses on small sky areas with high-resolution halo substructure.

Our simulation contains the discretization of the sky using the \textsc{HEALPix} scheme \citep{healpix}, which partitions the sky into equal-area pixels determined by the resolution parameter \( n_{\text{side}} \). The total number of pixels covering the sky is given by \( N_{\text{pix}} = 12 \times n_{\text{side}}^2 \). For example, with \( n_{\text{side}} = 2048 \), the sky is divided into approximately 50 million pixels. This results in an angular resolution on the order of arcminutes, allowing us to produce detailed maps of \( \kappa \) and shear \( \gamma \).

The entire sky covers an area of approximately \( 41,253 \) square degrees. With \( n_{\text{side}} = 2048 \), each pixel covers an area of about \( 0.00082 \) square degrees or \( 10,622 \) square arcseconds. Although our simulation provides a low-resolution view compared with our halo rendering method described in Section~\ref{sec:halo_rendering}, this method is sufficient to capture the essential features of the large-scale structure necessary for our analysis, especially providing a robust average \( \kappa \) and correlation of shear and convergence at intermediate to large scales beyond the effect of individual halos.

\subsubsection{Line-of-Sight Simulation\label{halos_method}}

While the large-scale structure simulation provides a smooth, full-sky representation, the line-of-sight simulation complements this by capturing the high-resolution, small-scale features that are critical for strong lensing analysis. In this section, we first perform halo rendering using the \textsc{Colossus} package to generate halos along light cones (see Section~\ref{sec:halo_rendering}). We then compute the lensing effects of these halos via ray-tracing (see Section~\ref{sec:ray_tracing}). Finally, non-linear line-of-sight corrections are applied to account for the varying contributions of halos located between the observer, deflector, and source (see Section~\ref{los_nonlinear_correction_method}).

\paragraph{Halo Rendering \label{sec:halo_rendering}}

The rendering of halos is a crucial component in our simulation, and its accuracy and precision are essential. To achieve this, we use the package \textsc{Colossus} \citep{diemer2018colossus}, which provides a framework for calculating the halo mass function. This mass function allows us to derive the number density of halos, enabling us to estimate the expected number of halos within a given light cone.

To ensure a realistic representation of the universe, we model the number of halos using a Poisson distribution, where the mean number is determined by the halo mass function. By applying growth functions, we assign redshifts and masses to the halos.

Our methodology involves several key steps:

\begin{enumerate}
    \item We compute the expected number of halos at different redshifts by combining the differential comoving volume with the halo number density.
    
    \item We generate a set of halo redshifts, where the number of halos at each redshift follows a Poisson distribution with the mean equal to the expected number. The redshift values are sampled from the cumulative distribution function (CDF) of the halo number density.
    
    \item For each halo at a given redshift, we assign a mass by randomly sampling from the halo mass function at that redshift.
\end{enumerate}

We use the \texttt{bhattacharya11} halo mass function model \citep{bhattacharya2011mass} implemented in the \textsc{Colossus} package.

The halo concentration affects the internal structure of halos and influences their lensing properties. We derive the halo concentration from the halo mass using the empirical fit presented in \citet{child2018halo}. Specifically, we use their Equation 19 and parameters from Table 2, which include data for both relaxed and unrelaxed halos. The concentration-mass relation is given by:

\begin{equation}
 c_{200 \rm c} = A(1+z)^d M^m 
 \end{equation}
 
where \( A \) is the pre-factor, \( d \) is the exponent for the redshift-dependent term, and \( m \) is the exponent for the mass term. We adopt \( A=75.4 \), \( d=-0.422 \), and \( m=-0.089 \).

In our study, halos are generated using a uniform distribution in position within the light cone at their assigned redshifts, providing a realistic depiction of their spatial distribution. This approach ensures that our simulations capture the inherent randomness and vastness of halo distributions in the universe. However, it should be emphasized that this uniform placement deliberately neglects halo angular clustering effects (the 2D spatial correlations of halo positions) and tidal alignment phenomena between halos and the large-scale structure (LSS).

\paragraph{Ray-Tracing Lensing Calculation \label{sec:ray_tracing}}

To calculate the lensing effects of individual halos, we employ \textsc{lenstronomy} \citep{birrer2018lenstronomy, birrer2021lenstronomy}. Given a halo characterized by its Navarro-Frenk-White (NFW) profile parameters (mass, concentration), \textsc{lenstronomy} computes all relevant lensing quantities, including convergence, shear, and deflection angles, with appropriate cosmological distances. 
Under the hood, \textsc{lenstronomy} utilizes a multi-plane lensing framework \citep[e.g.,][]{collett2016compound}, which generalizes the single-plane lens equation by sequentially applying deflections at each mass plane. For instance, in the case of \(\nu\) planes, the angular position \(\boldsymbol{x}_\nu\) on the \(\nu\)-th plane can be written as:

\begin{equation}
\boldsymbol{x}_\nu 
= \frac{D_{\nu}}{D_1} \boldsymbol{x}_1 
  - \sum_{\mu=1}^{\nu-1} \frac{D_{\mu\mu+1}}{D_\nu}\,\hat{\boldsymbol{\alpha}}_{\mu}\bigl(\boldsymbol{x}_{\mu}\bigr),
\end{equation}

where \(D_{\nu}\) is the angular diameter distance to plane \(\nu\), \(D_{\mu\mu+1}\) is the distance between planes \(\mu\) and \(\mu+1\), and \(\hat{\boldsymbol{\alpha}}_{\mu}(\boldsymbol{x}_{\mu})\) is the deflection angle on plane \(\mu\). 

After identifying the halos within the light cone and assigning their properties such as mass, redshift, and position, we proceed to compute the gravitational lensing effects they induce. To accurately capture the non-linear line-of-sight contributions, we decompose the total LOS effect into three distinct categories, based on the relative positions of the halos with respect to the observer, deflector, and source:
\begin{itemize}
    \item \textbf{Observer-Deflector (OD)}: Contribution from halos located between the observer and the deflector.
    \item \textbf{Observer-Source (OS)}: Contribution from halos along the line of sight between the observer and the source.
    \item \textbf{Deflector-Source (DS)}: Contribution from halos located between the deflector and the source.
\end{itemize}

For each of these configurations, we use the \textsc{lenstronomy} package to calculate the \( \kappa \) and \( \gamma \), which are second-order derivatives of the lensing potential. These calculated quantities are later combined using our non-linear correction formalism to properly account for the interactions among the different LOS contributions.

\paragraph{Non-linear Line-of-Sight Correction \label{los_nonlinear_correction_method}}

Structures along the LOS contribute additional convergence and shear that affect the observed image configurations. However, in strong lensing, due to the presence of a dominant deflector, objects along the LOS contribute differently to the total lensing effect depending on their relative positions with respect to the deflector. Perturbers located in front of the main deflector tend to exert a stronger influence on the lensing signal than those situated behind it (see Figure~5 of \citet{fleury2021line}). To account for these differences, we adopt the methodology of \citet{fleury2021line} and \citet{birrer2020tdcosmo}, which introduces non-linear line-of-sight corrections to the external convergence and shear.

The angular diameter distances in the presence of line-of-sight structures differ from those predicted by a homogeneous background cosmology. The corrected external convergence \( \kappa_{\text{ext}}^* \) is calculated using: 

\begin{equation}
    1 - \kappa_{\text{ext}}^* = \frac{(1 - \kappa_\text{OD})(1 - \kappa_\text{OS})}{1 - \kappa_\text{DS}}\,,
\end{equation}

where \( \kappa_\text{OD} \), \( \kappa_\text{OS} \), and \( \kappa_\text{DS} \) are the integrated convergences along the line of sight, modifying the distances from the observer to the deflector, the observer to the source, and the deflector to the source, respectively. This formulation accounts for the fact that the effective lensing contribution from background and foreground structures is modified by the tilt introduced by the main deflector.

The presence of multiple perturbers along the LOS also modifies the shear field. The corrected external shear \( \gamma_{\text{ext}}^* \) is computed as:

\begin{equation}
    \gamma_{\text{ext}}^* = \sqrt{[\gamma_\text{OD1} + \gamma_\text{OS1} - \gamma_\text{DS1}]^2 + [\gamma_\text{OD2} + \gamma_\text{OS2} - \gamma_\text{DS2}]^2},
\end{equation}

where \( \gamma_\text{OD1} \) and \( \gamma_\text{OD2} \) are the two components of shear for the Observer-Deflector configuration, and similarly for the other configurations.

These corrections account for the fact that the lensing effects from different halos along the LOS do not simply add linearly due to their interactions with each other and with the main deflector. These non-linear correction improve the accuracy of our lensing simulations by incorporating the effects of secondary lensing planes along the LOS. By incorporating these non-linear effects, we improve the accuracy of our lensing simulations and better represent the cumulative impact of all mass distributions between the observer and the source.

\subsection{Integration of Large-Scale Structure and Line-of-Sight Simulations \label{sec:method_Integration}}

The integration of Large-Scale Structure and Line-of-Sight Simulations using the \textsc{GLASS} \citep{GLASS} package from Section~\ref{LSS_method} and Section ~\ref{halos_method} is designed to capture the full complexity of the gravitational lensing signal by merging the strengths of each method.

The primary goal is to combine the shear maps, denoted by \( \gamma \), from both sources to obtain a comprehensive representation of the total shear, \( \gamma_{\text{tot}} \). We combine two complementary datasets:  

1. Smooth background shear from the \textsc{GLASS} simulation (LSS shear, \( \gamma_{\text{LSS}} \)), representing the average cosmic tidal forces.  

2. Fine-scale halo contributions (\( \gamma_{\text{halos}} \)), capturing perturbations from individual galaxies and clusters near the lens.

The total shear \( \gamma_{\text{tot}} \) is the vector sum of \( \gamma_{\text{LSS}} \) and \( \gamma_{\text{halos}} \). Although the large-scale structure and line-of-sight halos are inherently correlated, in our analysis we assume they are statistically independent, so that their shear orientations can be treated as uncorrelated and randomly distributed in angle. This allows us to approximate: 

\begin{equation}\label{eq:gamma_tot}
|\gamma_{\text{tot}}|^2 = |\gamma_{\text{halos}}|^2 + |\gamma_{\text{LSS}}|^2 + 2|\gamma_{\text{halos}}||\gamma_{\text{LSS}}| \cos(\Delta\theta),
\end{equation}

where \( \Delta\theta = \theta_{\text{LSS}} - \theta_{\text{halos}} \) is the difference in shear angles, where  $\theta_{\text{LSS}}$ and $\theta_{\text{halos}}$ are independent.

For the convergence, we combine the contributions from both sources:

\begin{equation}\label{eq:kappa_tot}
\kappa_{\text{tot}} = \underbrace{\kappa_{\text{LSS}}}_{\text{Large-scale background}} + \underbrace{\kappa_{\text{halos}} - \langle \kappa_{\text{halos}} \rangle}_{\text{Halo perturbations beyond the mean}}
\end{equation}

where \( \langle \kappa_{\text{halos}} \rangle \) is the average convergence due to the halos. This subtraction ensures that we avoid double-counting the mean density already included in the GLASS simulation.

\subsection{Evaluating Selection Effects via Simulation Pipeline\label{sec:bias_evaluating}}

Having obtained realistic maps that capture the line-of-sight effects through the external convergence and shear from \ref{sec:building the kg}, we are now positioned to test how these environmental factors influence the selection of strong lensing systems. 

\begin{figure*}
    \centering\includegraphics[width=\textwidth]{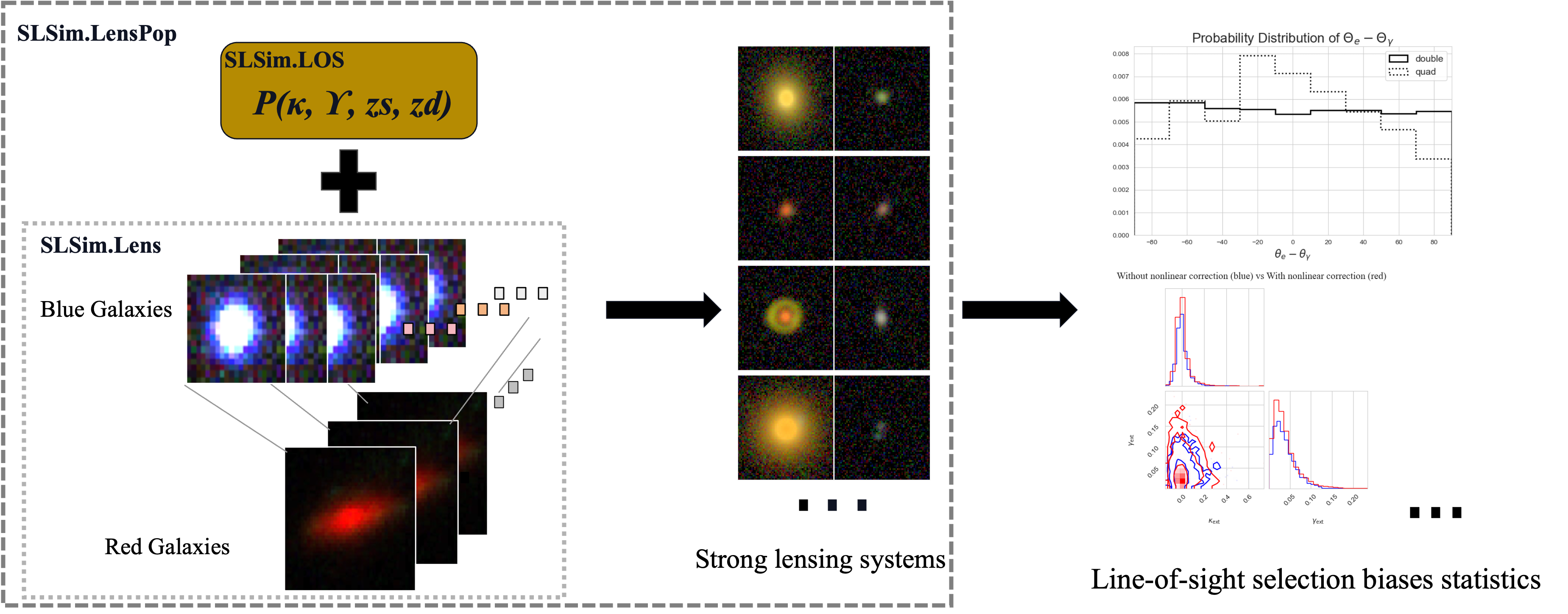}
    \caption{To assess how line-of-sight (LOS) effects influence lens detection and parameter inference, we incorporate our external convergence and shear distributions into \textsc{SLSim} \citep{slsim}. \textsc{SLSim} is a simulation framework designed to generate realistic populations of strong lenses under conditions similar to those of large-scale surveys (e.g., LSST). Using a Monte Carlo approach, it combines astrophysical models for source and deflector galaxies with observational criteria (including magnitude limits, image separations, and selection cuts) to produce lens samples. By comparing simulated lenses generated under different conditions (such as with and without LOS corrections), we quantify selection biases and refine the understanding of their impact on cosmological analyses.}
    \label{fig:slsim}
\end{figure*}

To assess the selection biases in strong gravitational lensing, we utilize the \textsc{SLSim} simulation tool \citep{slsim} with a Monte Carlo approach to generate catalogs of source and lensing galaxies, as illustrated in Figure~\ref{fig:slsim}. In this process, blue galaxies serve as sources, while red galaxies act as lenses.

\subsubsection{Galaxy Populations (Source and Deflector)}

We generate both source (blue) and deflector (red) galaxy populations using a combination of the \textsc{SLSim} \citep{slsim} simulation framework and \textsc{Skypy} \citep{skypy}, following the general methodologies provided. The source galaxies (late-type/blue) are characterized by various properties generated using the \textsc{Skypy} \citep{skypy} and \textsc{Astropy} \citep{astropy}, and have their redshifts, magnitudes, spectral coefficients, sizes, and ellipticities drawn from the Schechter luminosity function \citep{schechter1976analytic}. For deflector galaxies (early-type/red), which serve as lensing galaxies, we similarly employ \textsc{SLSim} and \textsc{Skypy} prescriptions for luminosity, size, and ellipticity, referencing standard velocity dispersion functions and scaling relations for early-type galaxies. Stellar masses are linked to velocity dispersions using empirical power-law relations \citep{auger2010sloan}. LSST filters calculated using \textsc{Speclitefilters} \citep{desihub_speclite} and applied to determine apparent magnitudes. In our baseline simulations each deflector galaxy’s mass distribution is represented by a singular isothermal ellipsoid (SIE) profile (density slope $\gamma=2$) \citep{SIE_paper} and the velocity dispersion $\sigma_v$ for each SIE is drawn from the SDSS stellar velocity‐dispersion function as described above, using the \textsc{SLSim} default distributions. The relationship between light and mass eccentricities was modeled based on the scaling and scatter parameters derived from data presented in Table 1 of \cite{sheu2024ProjectDinos}, incorporating a scaling factor of 1.135 and accounting for scatter in both eccentricities (0.127) and their orientation angles (0.319). Note that we do not include a separate NFW host‐halo component in the simulation galaxy‐scale lensing systems; instead, all halo effects enter exclusively through the external convergence ($\kappa_{\rm ext}$) and external shear ($\gamma_{\rm ext}$).

The galaxy sizes and ellipticities are assigned using the beta ellipticity distribution, as presented in III.A of \cite{kacprzak2020monte}, with distinct parameters for red and blue galaxies. The filters used for photometric properties are LSST \(g\), \(r\), \(i\), \(z\), and \(y\) bands, also shown in Table~\ref{tab:galaxy_properties}.

\begin{table*}
\centering
\begin{tabular}{|l|l|l|}
\hline
\textbf{Property} & \textbf{Blue Galaxies (Sources)} & \textbf{Red Galaxies (Lenses)} \\ \hline
\textbf{Redshift range} & \(z \in [0.0, 5.0]\) & \(z \in [0.0, 5.0]\) \\ \hline
\textbf{Luminosity function} & \(M^*(z) = -1.03z - 20.485\) & \(M^*(z) = -0.80z - 20.46\) \\
                             & \(\phi^*(z) = 0.0031 \times e^{-43.43z}\) & \(\phi^*(z) = 0.0028 \times e^{-1.06z}\) \\
                             & \(\alpha = -1.29\) & \(\alpha = -0.53\) \\ \hline
\textbf{Ellipticity} & \(e_{\text{ratio}} = 0.2, e_{\text{sum}} = 7.0\) & \(e_{\text{ratio}} = 0.45, e_{\text{sum}} = 3.5\) \\ \hline
\textbf{Filters} & LSST \(g\), \(r\), \(i\), \(z\), \(y\) bands & LSST \(g\), \(r\), \(i\), \(z\), \(y\) bands \\ \hline
\textbf{Sizes} & Derived from redshift \(z\) & Derived from redshift \(z\) \\ \hline
\end{tabular}
\caption{\label{tab:galaxy_properties}Key properties of blue (source) and red (lens) galaxy populations used in the simulations. Source galaxies are modeled as late-type (blue) galaxies, while lens galaxies are represented as early-type (red) galaxies. The table summarizes the redshift ranges, Schechter luminosity function parameters (\(M^*\), \(\phi^*\), \(\alpha\)), ellipticity distributions, photometric bands, and size dependencies for both populations.}
\end{table*}

\subsubsection{Incorporation of AGN as Background Sources}
To explore LOS effects on point-like, variable sources, we extend our pipeline to include active galactic nuclei (AGN) by using the \texttt{PointSources} class with the \texttt{quasar} type from \textsc{SLSim} \citep{slsim}.  AGN, being effectively unresolved and intrinsically variable, are ideal for time‐delay cosmography. The quasar population is generated using the \texttt{QuasarRate} class, which implements the quasar luminosity function from Richards et al. (2006). By default, this class adopts a double power-law luminosity function with parameters, with best-fit values from the SDSS DR3 survey \citep{SDSS_DR3, qusar_pop}. Redshifts are sampled from the comoving number density up to z=5.0. For each quasar, an absolute magnitude is drawn from the luminosity function at its redshift and converted to an apparent i-band magnitude, incorporating distance modulus and K-corrections derived from \citet{qusar_pop}.

\subsubsection{Cosmology prior} 

Unless specified otherwise, all simulations assume a flat-\(\Lambda\)CDM cosmology with \(H_0 = 70\) km s\(^{-1}\) Mpc\(^{-1}\) and \(\Omega_{\rm m} = 0.3\). The LOS halos are generated using the \textsc{GLASS} \citep{GLASS} large-scale structure implementation, galaxy populations are modeled using the \textsc{SLSim} \citep{slsim} tool, and the final strong lensing calculations are performed under these assumptions.
 
Additional settings and variations in cosmological parameters are explored in subsequent sections, as shown in Section~\ref{sec:results_different_cosmology}. We compare two observationally motivated cosmologies with different $\sigma_8$ and  $\Omega_{\rm m}$, enabling quantitative tests of cosmology-dependent selection effects. Users can similarly implement alternative cosmologies by modifying \textsc{GLASS} and \textsc{SLSim} configuration files for building external convergence and external shear distributions under different cosmologies. Furthermore, these alternative settings can be directly tested in \textsc{SLSim}, allowing users to assess how different cosmological conditions affect the properties of strong lensing events.

\subsubsection{Strong Lensing Selection Criteria}

We establish a strong lensing selection criteria that are applied after randomly pairing source and deflector galaxies and incorporating the full line-of-sight effects. The \textsc{SLSim} tool assesses the plausibility of strong lensing configurations based on a set of criteria:

\begin{itemize}
    \item The lens galaxy must be at a lower redshift than the source galaxy.
    \item The angular Einstein radius must fall within the predefined image separation limit of at least \(1.0\) arcseconds.
    \item The lensing galaxy must produce at least two images of the source.
    \item For extended sources, the magnified brightness of the lensed arc must exceed the observational threshold of magnitude 28 in the \(g\)-band.
\end{itemize}

\subsubsection{Simulation Procedure \label{sec:method：bias_simulation}}

We begin by randomly positioning galaxies and then validate potential strong lensing candidates against the established criteria. In our galaxy-galaxy catalog, we incorporate our precomputed redshift-dependent joint distributions of external convergence and external shear into the simulations. This ensures that only realistic and observable lensing configurations contribute to our assessment of selection biases.

Additionally, we compute the effects of the line of sight for different cosmological models and compare them with each other.

\section{Results}

\subsection{Joint Distributions of External Convergence and External Shear \label{sec:results_joint_distributions}}

\begin{figure}
    \centering
    \includegraphics[width=0.4\linewidth]{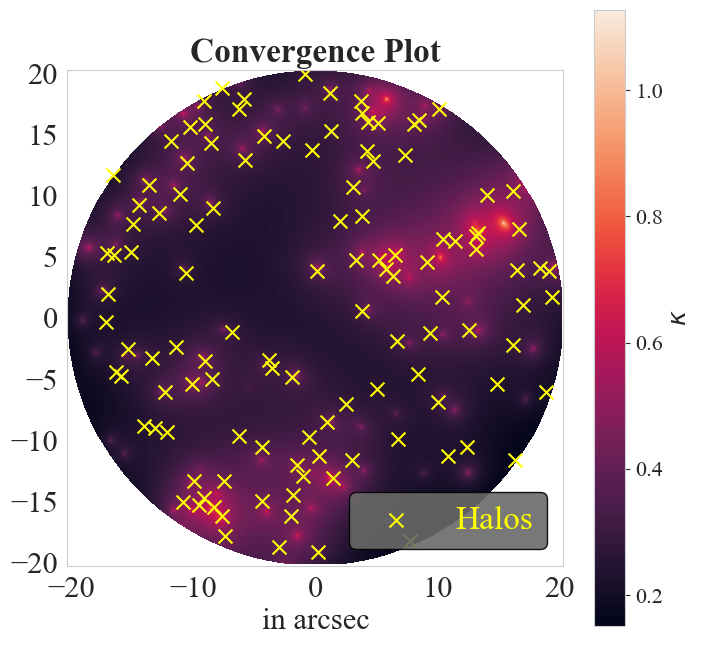}
    \caption{Convergence map showing the distribution of $\kappa$ within a radius of $\simeq 20$ arcseconds from one realization of halos rendering. The yellow crosses represent the positions of halos in the lensing model, with coordinates provided in arcseconds relative to the center. Due to lensing distortions from foreground halos affecting the observed convergence of background halos, the peaks in the convergence map do not necessarily coincide with the exact positions of the halos. The halos are modeled with masses ranging from $2.0 \times 10^{11} \ M_\odot$ to $1.0 \times 10^{16} \ M_\odot$ and are distributed up to a maximum redshift of $z=5.0$.}
    \label{fig:one halos rendering}
\end{figure}

Following the methodology described in Section~\ref{sec:building the kg}, we first generated the external convergence ($\kappa_{\text{ext}}$) and external shear ($\gamma_{\text{ext}}$) with halos rendering. Each halos rendering will give us a joint ($\kappa_{\text{ext}}$, $\gamma_{\text{ext}}$), with one of the rendering shown in Figure~\ref{fig:one halos rendering}. After halos rendering, we combine the LOS distributions from halos rendering with distributions from large scale structure using Equation~\ref{eq:gamma_tot} and Equation~\ref{eq:kappa_tot}.

We calculated the joint distributions of external convergence ($\kappa_{\text{ext}}$) and external shear ($\gamma_{\text{ext}}$), both with and without applying the non-linear line-of-sight corrections. The resulting data were stored in HDF5 format \citep{HDF5HDF5HDF5} for efficient access and analysis.

\begin{figure}
\centering
\begin{subfigure}[b]{0.45\columnwidth}
    \centering
    \includegraphics[width=\linewidth]{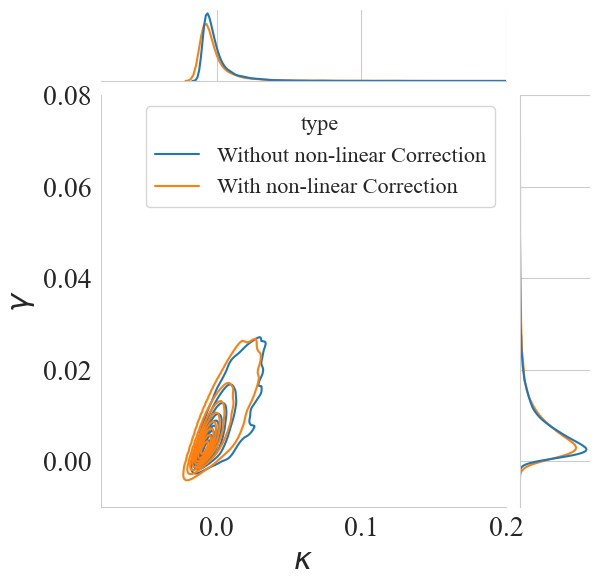}
    \caption{$z_{\rm s} = 1.0$, $z_{\rm l} = 0.5$}
    \label{fig:distri1}
\end{subfigure}
\hfill
\begin{subfigure}[b]{0.45\columnwidth}
    \centering
    \includegraphics[width=\linewidth]{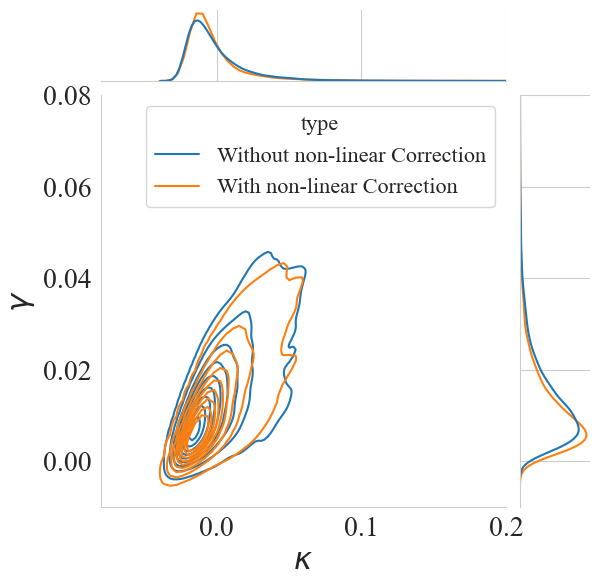}
    \caption{$z_{\rm s} = 1.5$, $z_{\rm l} =  0.5$}
    \label{fig:distri2}
\end{subfigure}

\vspace{0.3cm}

\begin{subfigure}[b]{0.45\columnwidth}
    \centering
    \includegraphics[width=\linewidth]{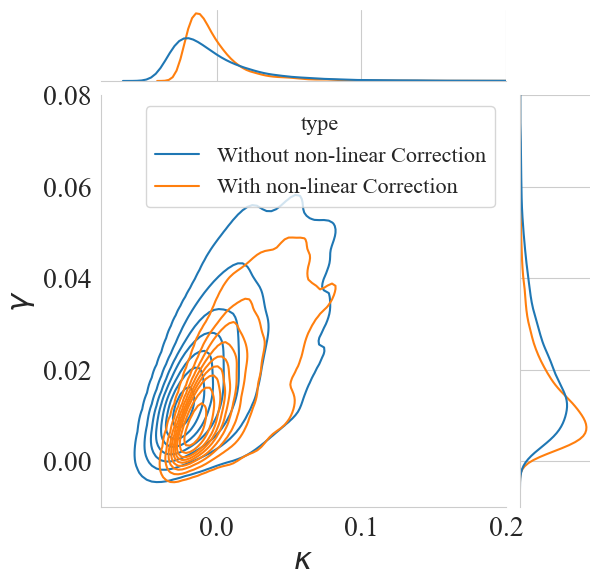}
    \caption{$z_{\rm s} = 2.0$, $z_{\rm l} = 0.5$}
    \label{fig:distri3}
\end{subfigure}
\hfill
\begin{subfigure}[b]{0.45\columnwidth}
    \centering
    \includegraphics[width=\linewidth]{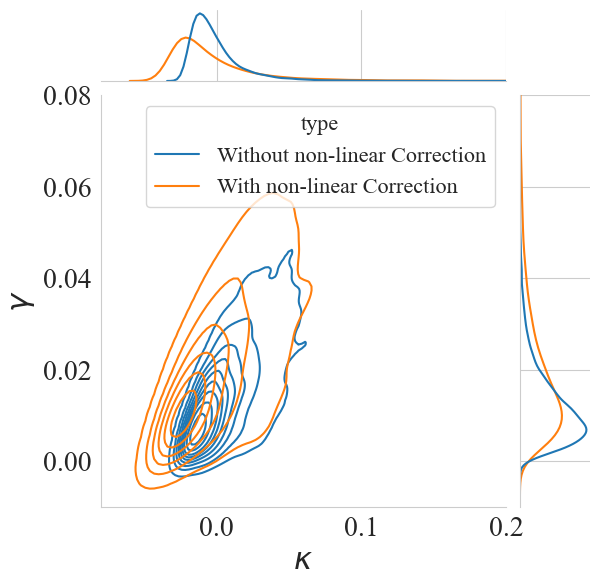}
    \caption{$z_{\rm s} =1.5$, $z_{\rm l} =  1.0$}
    \label{fig:distri4}
\end{subfigure}
\caption{The joint distributions of external convergence and external shear, with a minimal rendering halo mass of $10^{11} M_{\odot}$. The yellow contours are distributions with non-linear correction and the blue one are without the non-linear correction. As shown in figures, with the different source redshift ($z_{\rm s}$) but same major-deflector redshift ($z_{\rm l}$). Showing as the object between observer with major-deflector has larger lensing effect, causing higher source redshift plots have larger divergence for non-linear correction distributions. Additionally, after applying non-linear correction, the distributions exhibit a systematic divergence, with \(\kappa_{\text{ext}}\) generally increasing and shear becoming smaller at higher \(z_s\).}
\label{fig:quad}
\end{figure}

Figure~\ref{fig:quad} presents these joint distributions for a minimal halo rendering mass of $10^{11}\ M_\odot$. In the plots, the yellow contours represent the distributions obtained with the non-linear correction, while the blue contours correspond to the distributions without the non-linear correction. The figures illustrate cases with different source redshifts but the same deflector redshift.

From the figures, we observe that when the source redshift increases while keeping the deflector redshift constant, the divergence (quantified as standard deviation) between the distributions with and without non-linear correction become from small to larger. This is because objects located between the observer and the deflector contribute more significantly to the lensing effect at higher source redshifts, leading to greater discrepancies when non-linear correction are included.


The data used to compute these distributions, including various settings (such as $\sigma_8 = 0.810$ and $\sigma_8 = 0.825$, with or without the non-linear correction, etc) are publicly available.\footnote{The Line of Sight External Convergence and Shear Distributions Files: \href{https://github.com/LSST-strong-lensing/data_public/tree/main/Line_of_sight_kg_distributions}{https://github.com/LSST-strong-lensing/data\_public/tree/main/Line\_of\_sight\_kg\_distributions}}.

\subsection{Distributions with and without Non-linear Correction}

As discussed in Section~\ref{los_nonlinear_correction_method}, the non-linear correction accounts for the fact that objects along the line of sight contribute differently to the overall lensing signal depending on their relative positions to a dominant deflector. As the joint distributions of $\kappa_{\text{ext}}$ and $\gamma_{\text{ext}}$ with or without the non-linear correction presented in Figure~\ref{fig:quad} and Section~\ref{sec:results_joint_distributions}. After incorporating these effects into the \textsc{SLSim} simulation pipeline, we can test how the line-of-sight selection affects the final strong lensing configurations.

\begin{figure*}
    \centering
    \includegraphics[width=1.0\textwidth]{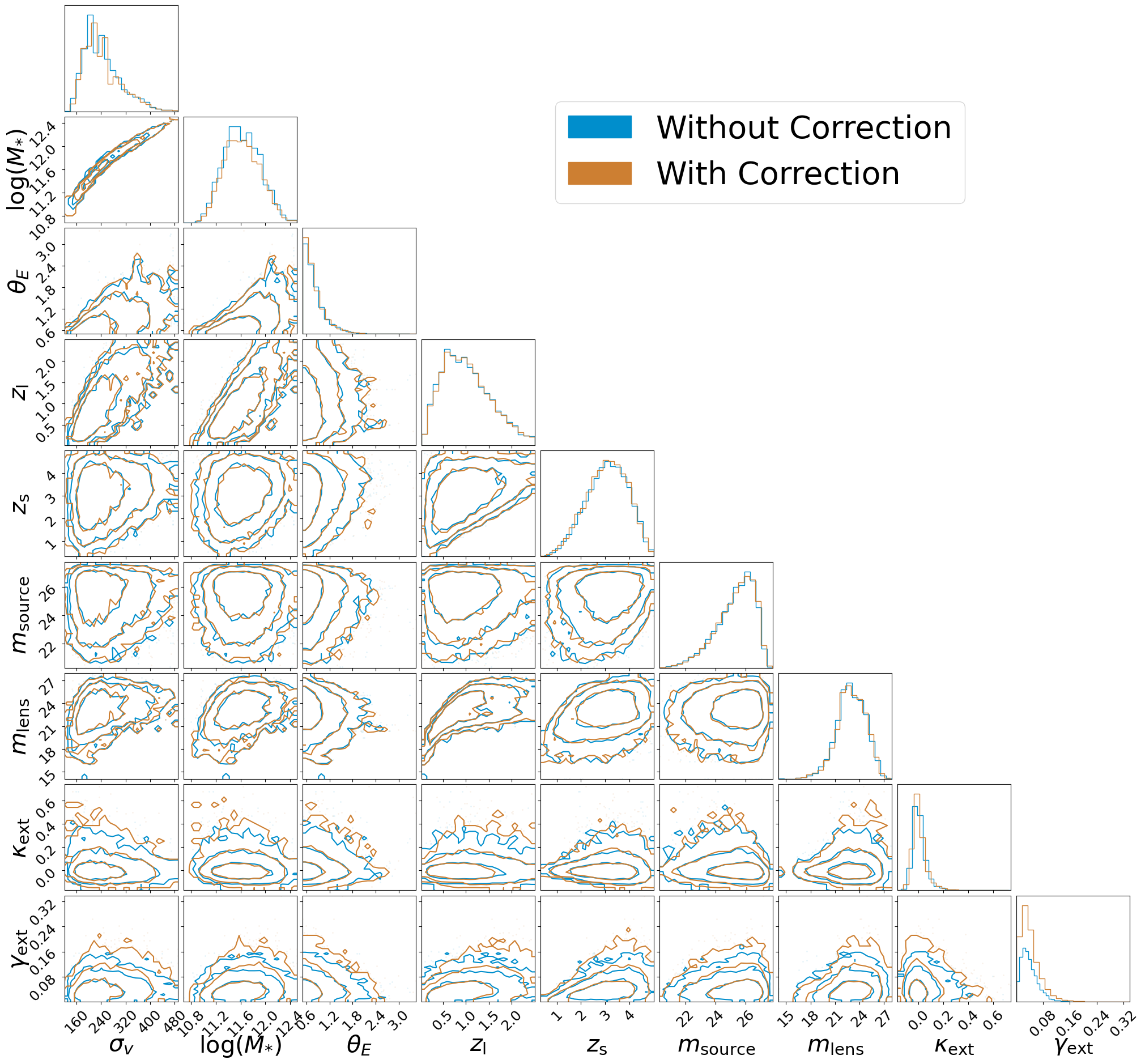}\caption{Posterior distribution of parameters of strongly lensed galaxies with accessing line of sight effect. (Orange are with the non-linear correction; blue are galaxies without non-linear correction). Parameters include velocity dispersion ($\sigma_v$, in km/s), logarithmic deflector stellar mass ($\log(M_{*})$, in $\log(M_{*}/M_{\odot}$), Einstein radius ($\theta_E$, in arcseconds), lens galaxy redshift ($z_{\rm l}$), source galaxy redshift ($z_{\rm s}$), apparent magnitude of the lensed source ($m_{\rm source}$), apparent magnitude of the lens galaxy ($m_{\rm lens}$), external convergence ($\kappa_{\rm ext}$), and external shear ($\gamma_{\rm ext}$).}
    \label{fig:cornor_wwolos}
\end{figure*}

In Figure~\ref{fig:cornor_wwolos}, we present a corner plot of various physical parameters of strongly lensed galaxies, comparing cases with and without the non-linear line-of-sight corrections.

From the Figure~\ref{fig:cornor_wwolos}, it is evident that including the non-linear correction leads to slightly larger divergence of external convergence and shear than without the correction. This indicates that non-linear line-of-sight effects enhance the lensing distortions and should be considered for accurate modeling of galaxy-galaxy lensing.

\subsection{Impact of Different Cosmological Models \label{sec:results_different_cosmology}}

We applied our methodology as described in Secction~\ref{sec:method：bias_simulation} to test the effects of different cosmological parameters on the line-of-sight lensing effects. Specifically, we considered the \textit{Planck}-2018 cosmology with $\sigma_8 = 0.810$ and $\Omega_\text{m} = 0.3111$ \citep{aghanim2020planck}, and the combined DES+KiDS cosmology with $\sigma_8 = 0.825$ and $\Omega_\text{m} = 0.280$ \citep{abbott2023y3+}. We incorporated these cosmologies into both our large-scale simulations and halo rendering methods. Subsequently, these LOS distributions are fed into our strong lensing simulation pipeline (\textsc{SLSim}) under the corresponding cosmological setting, where we test the impact of the LOS effects on the observable properties of strongly lensed galaxies.

\begin{figure*}
    \centering
    \includegraphics[width=1.0\textwidth]{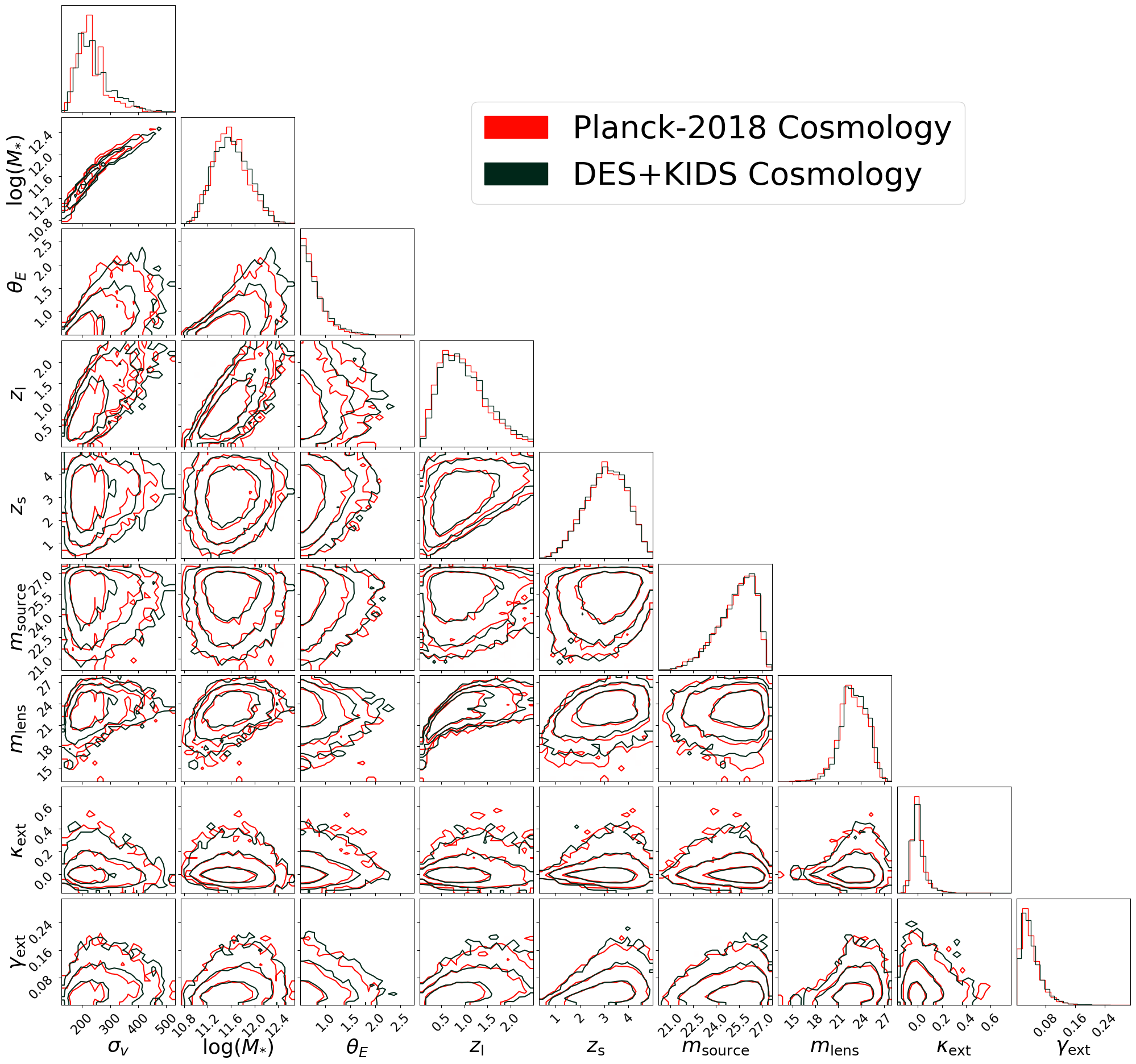}
    \caption{The Corner plot of strongly lensed galaxies’s red are with the \textit{Planck}-2018 cosmology; dark green are DES+KIDS cosmology). Parameters are same as Figure~\ref{fig:cornor_wwolos}.}
    \label{fig:different_cosmo}
\end{figure*}

Figure~\ref{fig:different_cosmo} displays a corner plot for strongly lensed galaxies under these two cosmological models. The red contours represent the \textit{Planck}-2018 cosmology, while the dark green contours correspond to the DES+KiDS cosmology.

While our analysis indicates no statistically significant difference in the overall line-of-sight effects between the two cosmologies, we find a notable variation in the number of strongly lensed cases.

\subsection{Selection of Quad and Double Lensing Images}

\begin{figure}
    \centering
    \includegraphics[width=1\linewidth]{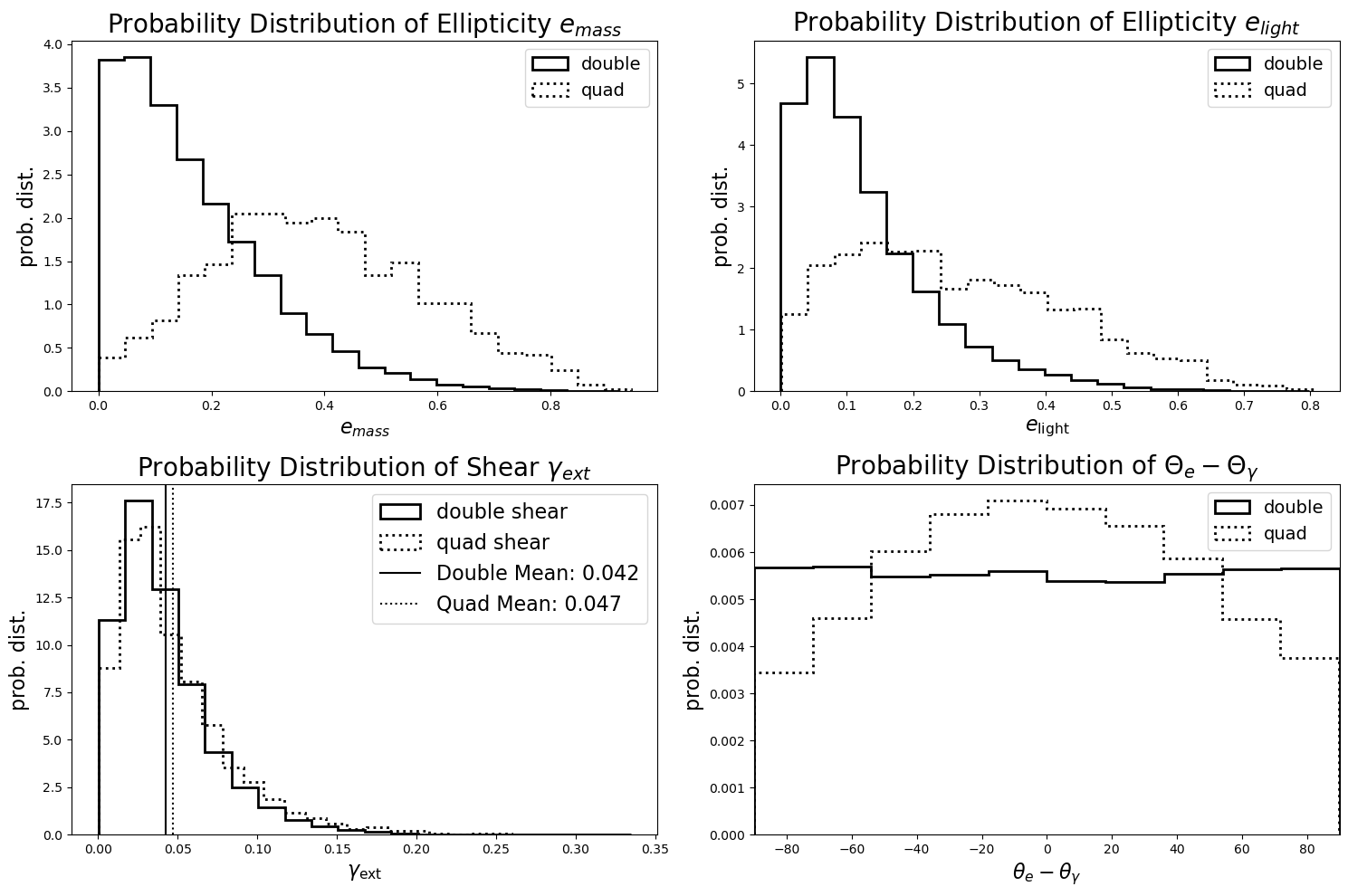}
    \caption{Probability distribution of the lens galaxy properties for double (solid) and quad (dotted) lenses, derived from the mock lens catalogue. \textbf{Top-left panel}: Probability distribution of the mass ellipticity, \( e_{\text{mass}} \), indicating the lens galaxy shape defined by mass distribution for double and quad lenses. \textbf{Top-right panel}: Probability distribution of the light ellipticity, \( e_{\text{light}} \), representing the observed shape from light distribution in double and quad lenses. \textbf{Bottom-left panel}: Probability distribution of external shear, \( \gamma_{\text{ext}} \), comparing the shear environments for double and quad lenses. Mean shear values are marked for each lens type, highlighting typical environmental shear strengths. \textbf{Bottom-right panel}: Distribution of the difference in position angles between the lens galaxy's ellipticity axis (\( \phi_e \)) and the direction of external shear (\( \phi_{\gamma} \)). Under the assumption that both \( \phi_e \) and \( \phi_{\gamma} \) are uniformly distributed, the difference \( \phi_e - \phi_{\gamma} \) is expected to approximate a uniform distribution across the population.
}
    \label{fig:double_quad}
\end{figure}

We further investigate the lensing biases in our mock lens catalog by examining how the lens population differs from a typical galaxy ensemble. Two key aspects are highlighted: the distribution of galaxy ellipticity (top panels of Figure~\ref{fig:double_quad}) and the alignment between galaxy orientation and external shear (bottom-right panel of Figure~\ref{fig:double_quad}).

The top panels of Figure~\ref{fig:double_quad} show the probability distribution of the lens galaxy ellipticity \( e \) for double (solid lines) and quadruple (dotted lines) lenses, as obtained from our mock lens catalog. The top-left panel represents the mass ellipticity (\( e_{\text{mass}} \)), and the top-right panel shows the light ellipticity (\( e_{\text{light}} \)). The distributions indicate that quad lenses tend to have higher ellipticities compared to double lenses, consistent with the expectation from lensing theory that higher ellipticity increases the likelihood of producing quad images. These are are are consistent with the expectation from lensing theory that higher ellipticity increases the likelihood of producing quad images \citep{oguri2010gravitationally}.

The bottom panels of Figure~\ref{fig:double_quad} analyze the alignment between the lens galaxy’s orientation and external shear. Our results show a clear bias: in quad lenses, the external shear tends to align with the galaxy’s major axis, while in double lenses, it is more likely to align with the minor axis. This is consistent with \citep{keeton1997shear, oguri2010gravitationally}.

Also, we calculated the ratio of quadruple to double lenses in our simulated strongly lensed cases:
\[
\frac{\text{Quad}}{\text{Double}} = \left\{
\begin{aligned}
&\sim 4.61\% \quad \text{(without line-of-sight)} \\
&\sim 5.13\% \quad \text{(LOS without non-linear correction)} \\
&\sim 5.24\% \quad \text{(LOS with non-linear correction)}
\end{aligned}
\right.
\]
Although our estimated quad-to-double ratio is relatively small (typically expected to be around $15\ \text{to}\ 25\%$), it is comparable in magnitude to the values found for the "Mixed" and "Unbiased" cases in Table 2 of \citet{mandelbaum2009galaxy}. It is important to note that the absolute quad-to-double ratio is highly sensitive to the underlying galaxy population assumptions, particularly the choice of mass ellipticity in galaxy simulations (also determined in Table 2 of \citet{mandelbaum2009galaxy}). In our case, we focus primarily on the relative changes in the quad-to-double ratio under different line-of-sight conditions rather than the absolute value.

These results indicate that the inclusion of line-of-sight effects and non-linear correction has a small impact on the quad-to-double lens ratio in our simulations.  However, neglecting external convergence and shear introduces a slightly more significant effect, reducing the quad-to-double ratio. Nevertheless, this effect remains less significant when compared to the influence of ellipticity. Therefore, while line-of-sight corrections and external factors play a role, the intrinsic ellipticity of the deflector has a more substantial impact in our simulations.

\subsection{Impact of LOS convergence on \(H_0\) inference}

The estimation of the Hubble constant \( H_0 \) from time-delay measurements in strong gravitational lensing is sensitive to the mass distribution along the LOS. The external convergence (\( \kappa_{\text{ext}} \)) contributes an additional lensing effect that can bias the inferred value of \( H_0 \) if not properly accounted for. The relationship between the inferred Hubble constant (\( H_0^{\text{o}} \)) (without accounting for the $\kappa_\text{ext}$) and the true value is given by:
\begin{equation}
    H_0 = (1 - \kappa_{\text{ext}}) H_0^{\text{o}}.
\end{equation}

\begin{figure}
    \centering
    \includegraphics[width=1.0\linewidth]{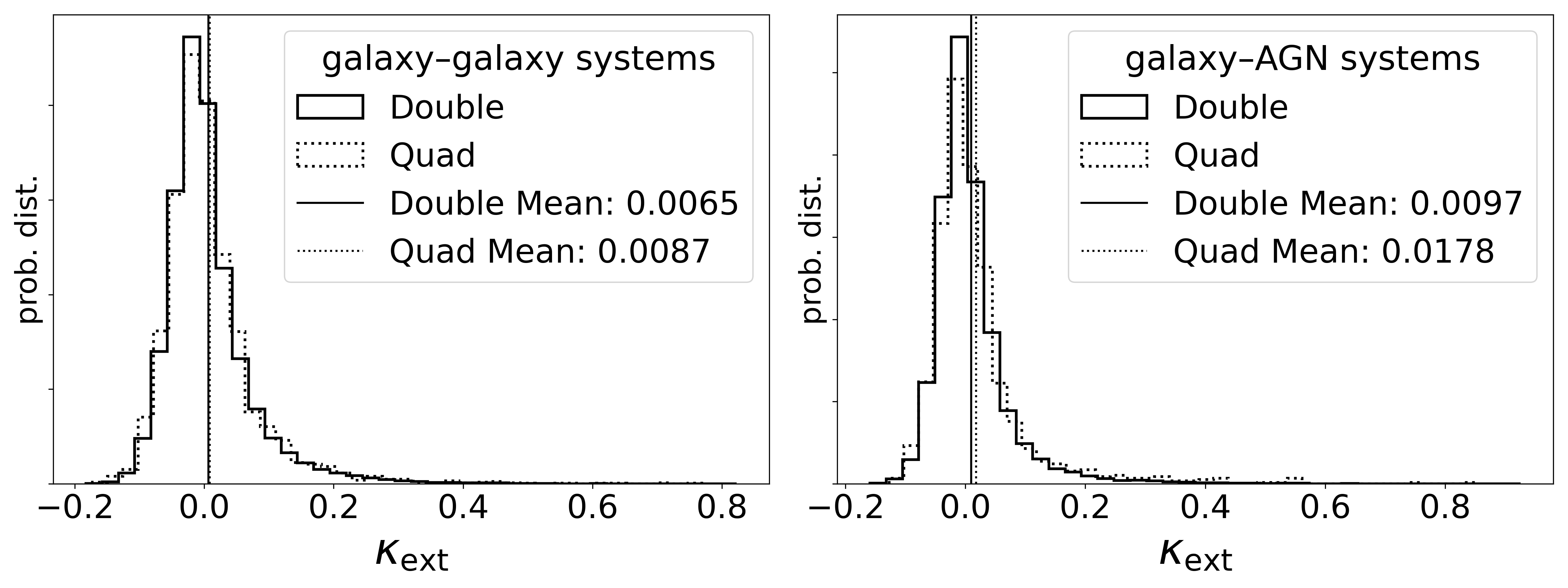}
    \caption{Probability distribution of external convergence (\(\kappa_{\text{ext}}\)) for double-image (solid black line) and quadruple-image (dotted black line) lens systems with the non-linear correction applied. Left: galaxy–galaxy; Right: galaxy–AGN. Vertical lines mark the means \(\langle \kappa_{\text{ext}}\rangle\): galaxy–galaxy doubles \(0.00649\), quads \(0.00872\); galaxy–AGN doubles \(0.00975\), quads \(0.01784\).
    }
    \label{fig:external_convergence}
\end{figure}

Historically, this bias has been addressed in precision analyses since \cite{suyu2010dissecting}, with \( \kappa_{\text{ext}} \) estimated using observational proxies or simulations. Our approach builds on these methods, incorporating non-linear correction to account for interactions between LOS structures and the primary lens, enhancing the accuracy beyond first-order (linear) treatments. To quantify the impact of line-of-sight structures on \( H_0 \), we computed the mean external convergence for both double-image and quadruple-image lens systems using our simulated data. The results are as follows and also as shown in Figure~\ref{fig:external_convergence}.

Using our simulations, we measure the mean \(\kappa_{\rm ext}\) for both galaxy–galaxy and galaxy–AGN lenses. We present a two-part analysis to quantify:
1. the total impact of LOS convergence on \( H_0 \) if entirely uncorrected; and  
2. the marginal bias introduced when LOS effects are modeled neglecting interactions between LOS structures and the primary lens.

\textbf{Total LOS-Induced Bias in \( H_0 \):} These mean values from Figure~\ref{fig:external_convergence} indicate that, on average, external convergence introduces a bias, if the effect of external convergence is not accounted for, galaxy–galaxy systems bias \(H_0\) by \(\sim0.65\%\) (doubles) and \(\sim0.87\%\) (quads), \(\sim0.66\%\) (in general); while galaxy–AGN systems bias by \(\sim0.97\%\) (doubles) and \(\sim1.78\%\) (quads), \(\sim1.02\%\) (in general).

The larger bias observed in quadruple-image lenses can be attributed to their increased sensitivity to the surrounding mass distribution and line-of-sight structures. Quadruple lenses often reside in more complex environments or are more affected by intervening matter, leading to higher external convergence values compared to double lenses.

Although the bias percentages may appear small, they are significant in the context of precision cosmology. Current measurements of \( H_0 \) aim for uncertainties below 2\%, and a systematic bias of over 1\% due to unaccounted external convergence could contribute substantially to discrepancies between different \( H_0 \) measurement methods. 
The impact becomes even more significant in smaller samples. This is evident from the 68\% central intervals reported in Table~\ref{tab:bias_ranges}: when LOS effect are not applied, the spread of the bias widens noticeably. Note that current analyses do incorporate corrections for external convergence \citep[e.g.,][]{greene2013, wells2023tdcosmo}.

\begin{table}
\centering
\caption{68\% central range of the $H_0$ bias from neglecting LOS effect of lensing systems.}
\label{tab:bias_ranges}
\begin{tabular}{llcc}
\hline
System            & Image   & With correction             & Without correction          \\
\hline
Galaxy--galaxy     & Double  & $[-4.59\%,\ 5.26\%]$      & $[-4.61\%,\ 5.19\%]$      \\
Galaxy--galaxy     & Quad    & $[-4.75\%,\ 5.51\%]$      & $[-4.81\%,\ 6.17\%]$      \\
Galaxy--AGN        & Double  & $[-3.70\%,\ 4.77\%]$      & $[-3.86\%,\ 5.49\%]$      \\
Galaxy--AGN        & Quad    & $[-4.09\%,\ 5.74\%]$      & $[-4.12\%,\ 6.21\%]$      \\
\hline
\end{tabular}
\end{table}

\textbf{ Bias from Neglecting Non-Linear LOS Corrections:} To isolate the effect of non-linear correction, we compared the inferred mean convergence values calculated with and without the non-linear treatment. We find that the additional bias incurred by using linear-only LOS models is:
\begin{align*}
    \Delta \kappa_{\text{ext}} &= 0.0003 \quad \text{for galaxy--galaxy systems}, \\
    \Delta \kappa_{\text{ext}} &= 0.0010 \quad \text{for galaxy--AGN systems}.
\end{align*}

These translate into $ H_0 $ biases of approximately 0.03\% for galaxy--galaxy systems and 0.1\% for galaxy--AGN systems, respectively, relative to what one would obtain using current methods that omit non-linear correction. While these mean offsets are small, they represent systematic shifts that remain in population-level measurements and are particularly relevant for precision cosmology. 

Crucially, these values reflect only the average behavior—if the sample size is limited, the population mean may not be well constrained, and the resulting systematic bias in $H_0$ can be significantly larger. To illustrate this point, we report the 68\% central intervals for the inferred $H_0$ bias in Table~\ref{tab:bias_ranges}, comparing models with and without the non-linear correction.

Our findings underscore the necessity of incorporating accurate line-of-sight convergence corrections when using strong lensing systems to infer cosmological parameters. Neglecting the external convergence can lead to biased estimates of \( H_0 \), particularly in samples dominated by quadruple-image lenses. Therefore, detailed environmental studies and corrections for line-of-sight effects are essential for reducing systematic errors and enhancing the reliability of \( H_0 \) measurements from gravitational lensing.

\section{Discussion}

Our study presents an approach to modeling the LOS effects in strong gravitational lensing by constructing realistic joint distributions of external convergence (\(\kappa_{\text{ext}}\)) and external shear (\(\gamma_{\text{ext}}\)) to a sample-level analysis. By aggregating large-scale structure simulations with high-resolution halo renderings and applying non-linear correction, we have developed a method that can capture both the smooth background matter distribution and the perturbations from individual halos along the LOS. This methodology can help increase the precision of lensing efficiency estimations and provides a quantitative characterization of the effect of LOS structure on key observables in strong lensing studies, such as image distortions, time delays, and magnification patterns.

Our work builds upon and extends from previous studies that have investigated LOS effects in strong lensing. 
The influence of LOS structures on strong lensing selection effects has been highlighted in several works \citep{sonnenfeld2023selection}. For example, \cite{sonnenfeld2023selection} investigated how the distribution of sources and deflectors in lensing events differs from their parent populations. They emphasized the importance of understanding selection biases when studying subsets of the lens population. Our work contributes to this area by providing realistic joint distributions of \(\kappa_{\text{ext}}\) and \(\gamma_{\text{ext}}\), enabling a more accurate assessment of selection effects due to LOS structures.

\citet{fleury2021line} emphasized the necessity of considering both strong and weak lensing simultaneously when accounting for LOS effects. They introduced the tidal approximation to incorporate the influence of inhomogeneities along the LOS as external convergence and shear. Our approach implements the very same corrections that are presented by \citet{fleury2021line} from the non-linear coupling of the deflector with the LOS distortions.

\citet{oguri2010gravitationally} and \citet{collett2016observational} employed Monte Carlo simulations to model LOS effects by assigning probability distributions for external shear and convergence, deriving selection functions based on lensing cross-sections and image configurations. Our approach is very similar, as our simulations are also a result of a Monte Carlo simulation with very similar assumptions on the deflector and the source populations, produced by \textsc{SLSim}. Our method advances this by incorporating more accurate predictions for the LOS structure based on large scale structure and high-resolution halo models. This allows for a more detailed and accurate representation of LOS effects, especially in dense environments where individual halos can have significant impacts.

\citet{collett2016observational} used ray-tracing through cosmological simulations to derive joint distributions of external convergence and shear, exploring how differences in these quantities affect the observed properties of quadruple and double image configurations. Our approach differs by aggregating both large-scale and small-scale structures with non-linear correction, crucial for capturing contributions from individual halos along the LOS, particularly those near the dominant lens. 
Our approach has the flexibility to change the cosmology, can account for the non-linear coupling corrections \citep{fleury2021line} is inherently not resolution limited, and does not require large simulation boxes.

Our findings underscore the importance of accurately modeling LOS effects for precise measurements of the Hubble constant (\(H_0\)) and other cosmological parameters. As we apply the non-linear LOS corrections accounting for the fact that, in strong lensing systems, the presence of a dominant deflector alters the apparent positions and lensing contributions of structures along the LOS. We show that neglecting these non-linear effects introduces residual biases in population-level $H_0$ estimates—approximately 0.03\% for galaxy–galaxy systems and 0.1\% for galaxy–AGN systems. While these shifts show the non-linear LOS correction is important in precision cosmology, they can become even more significant in smaller samples or biased selections. Additionally, our analysis was conducted with an image separation cut of at least 1.0 arcseconds to ensure robust lens model selection and observability. Varying this separation threshold would alter the population characteristics, including LOS convergence distributions. As highlighted by \citet{li2021impact}, they demonstrated that unaccounted LOS structures could introduce systematic uncertainties in time-delay cosmography. By providing a method that accounts for these non-linear effects, our work enhances the reliability of cosmological inferences drawn from strong lensing observations.

Furthermore, our results have implications for the ratio of quadruple to double image configurations in strong lensing systems. We find that including non-linear LOS corrections leads to a slight increase in the quad-to-double lens ratio in our simulations. This suggests that accurate modeling of LOS structures is essential for understanding the observed distribution of lensing configurations and for reducing biases in statistical analyses of lensing populations. While previous studies \citep[e.g.,][]{keeton1997shear} found that compensatory external shear (e.g., unmodeled angular structure in the primary lens) minimally affects the quad/double ratio, our work highlights a distinct physical regime: environmental shear from large-scale structure perturbs the lens potential in a way that systematically alters image multiplicity statistics.

The method of explicitly rendering halo density profiles along the LOS enables us also to assess higher-order flexion effects on the lenses and the modeling thereof. We leave this to future work but highlight the potential need for it as described by e.g., \citep{McCully:2017, Duboscq:2024}.

Our model adopts a simplified uniform spatial distribution of halos, neglecting both their intrinsic angular clustering and potential shear orientation alignment with the LSS. While this approach avoids overcounting shear power by treating \( \gamma_{\text{halos}} \) and \( \gamma_{\text{LSS}} \) as independent terms, it omits two physical effects: (1) halo spatial clustering, which enhances non-linear lensing correlations, and (2) tidal alignment between halos and the LSS field that could coherently orient their shear angles. These simplifications provide a conservative baseline for isolating halo-scale effects, but future work will require incorporating halo clustering (e.g., via HOD modeling) and self-consistent shear-LSS alignment in high-resolution simulations to quantify their combined impact on lensing observables.

A complementary approach to investigate and correct-for LOS selection effects in a specific strong lens sample is to constrain the LOS around specific lenses based on galaxy number density proxies from photometric and spectroscopic data in the environment of the lens \citep[e.g.][]{Suyu:2010, Fassnacht:2011, Greene:2013, Rusu:2017} or LSS weak lensing convergence constraints \citep[e.g.,][]{Tihhonova:2020, Kuhn:2021}. Since the amount of information for a specific line of sight is limited, any of such methods will rely on priors on the parent distribution of shear and convergence. \cite{Wells:2024} demonstrated that a hierarchical analysis can be performed to estimate the sample-level bias for a population of 25 strong lenses.

The information compression with summary statistics to select analogue line of sights in numerical simulations is non-optimal and never includes the full context of the specific LOS in question.
Bayesian graph neural network developed by \citet{park2023hierarchical} offer alternative approaches for estimating external convergence from photometric data. These simulation-based inference methods inherently rely on training data that encode assumptions about the statistical properties of line-of-sight (LOS) structure. While current implementations may not fully capture non-linear interactions along the LOS, their framework is flexible: re-training such models on simulations that include these non-linear correction (e.g., higher-resolution halo catalogs or full 3D potential integration as implemented in this work) could directly address this limitation. Similarly, summary statistics approaches like those used in TDCOSMO \citep{birrer2020tdcosmo} could be upgraded by re-evaluating \(\kappa_{\text{ext}}\) estimates on sightlines processed with improved non-linear correction models. 

The two complementary approaches of a) a first-principled approach forward modeling specific strong lensing populations in the full cosmological context, such as being pursued with this work, and b) empirically characterizing the sightlines of strong lenses with available data must eventually result in the same outcome for a given strong lensing population. A combination of an empirical and first-principled approach will be able to further calibrate either of the two approaches, or perhaps can even lead to cosmological constraints itself with the expected correlation of strong lenses and LSS \citep{Birrer:2018, Hogg:2023}. Populating the LOS halos with actual galaxies we leave for future work. Initial models to link galaxies and halos have already been developed \citep{Abe:2025}.

Our methodology is specifically designed to be applied to large lens samples of the upcoming large-scale surveys such as LSST, \textit{Euclid}, and the Nancy Grace Roman Space Telescope. With the anticipated increase in the number of strong lensing systems, it becomes more time-consuming to perform detailed individual analyses for each system. Our approach provides efficient means of accounting for LOS effects across the lensing population, in particular for homogeneous lens populations whose selection effects, such as the detection probability, can be characterized.

\section{Conclusions}
In this work, we presented a method to construct a statistically representative distribution of the line of sight shear and convergence for strong gravitational lenses, by aggregating large-scale structure simulations with high-resolution halo rendering, and taking into account the correction from the non-linear bending at the main deflector.
We integrated this method into an end-to-end strong lensing population simulation, \textsc{SLSim}, and are able to predict the expected line of sight selection effects of a specified strong lensing population. Our method allows for a more accurate capturing of the LOS selection effects in strong lenses compared to evaluating the weak lensing effects from the observer to the source alone, the currently still most used approach.

Our semi-analytic approach to describing LOS effects allows us to investigate dependencies on cosmology, in particular the clustering of matter. We do not observe significant differences in the LOS distribution of lenses when varying the cosmological model within current observational constraints, which suggests that our methodology is robust against moderate variations in cosmological parameters. 
This shows the applicability of our approach across different cosmological scenarios.

Our method to produce a realistic population of strong gravitational lenses, including an accurate description of the LOS effect, creates the ability to understand lensing selection effects from first principles, a valid alternative to a data-informed approach using galaxy over-density statistics along the line of sight of strong lenses. In the future, additional selections based on specific lens finding, follow-up decisions, and any other kind that is introduced in the process of the analysis of a large sample of lenses can also be incorporated and propagated through an end-to-end analysis.

\section*{Acknowledgments}
\begin{itemize}

\item \textbf{DESC/SLSC acknowledgements}: 
This paper has undergone internal review in the LSST Dark Energy Science Collaboration.  
Alessandro~Sonnenfeld and Phil~Marshall served as internal reviewers and provided invaluable, detailed feedback.  
We also thank Thomas~Collett, Lucia~Marchetti, Adam S.~Bolton, Patrick~Wells, and Nicolas~Tessore for helpful discussions, and Tansu~Daylan for coordinating efforts within the LSST Strong Lensing Science Collaboration (SLSC).  
Dillon~Brout coordinated comments in the DESC Time-Delay Working Group.
\item \textbf{Author contributions}:  
TZ contributed to the methodology, wrote the code, ran the simulations, performed the analysis, and drafted the manuscript.  
SB conceived the project, designed the methodology, supervised the work, and revised the manuscript.  
AJS provided scientific guidance and reviewed the manuscript. 
NK help implemented LOS modules in \textsc{SLSim}, validated the pipeline, and provided feedback on the manuscript.  

\item \textbf{Grants}:
TZ gratefully acknowledges the support of the Undergraduate Research and Creative Activities (URECA) Program at Stony Brook University.
SB acknowledges support by the Department of Physics \& Astronomy, Stony Brook University.
Support for this work was provided by NASA through the NASA Hubble Fellowship grant HST-HF2-51492 awarded to AJS by the Space Telescope Science Institute (STScI), which is operated by the Association of Universities for Research in Astronomy, Inc., for NASA, under contract NAS5-26555. AJS also received support from NASA through the STScI grants HST-GO-16773 and JWST-GO-2974. 
NK is supported by the DoE/DESC as well as the Schmidt Futures Foundation.

\item \textbf{Codes}:  
This work made extensive use of \textsc{SLSim}\,\citep{slsim}, \textsc{GLASS}\,\citep{GLASS}, \textsc{lenstronomy}\,\citep{birrer2021lenstronomy}, \textsc{Colossus}\,\citep{diemer2018colossus}, \textsc{healpix}\,\citep{healpix}, \textsc{skypy}\,\citep{skypy}, \textsc{Astropy}\,\citep{astropy},
\textsc{Speclitefilters} \citep{desihub_speclite}, 
and core \textsc{Python} packages including \texttt{NumPy}\,\citep{numpy}, \texttt{SciPy}\,\citep{scipy}, and \texttt{Matplotlib}\,\citep{matplotlib}.  All other software packages are cited in the text.
\end{itemize}

\section*{Data Availability}

The code used in this work is available in the publicly accessible repository \textsc{Slsim}\footnote{\url{https://github.com/LSST-strong-lensing/slsim}} \cite{slsim}. Specifically, the following resources are most relevant to the analyses presented:

\begin{itemize}
    \item A tutorial on external convergence and shear distributions\footnote{\url{https://github.com/LSST-strong-lensing/slsim/blob/main/notebooks/external_convergence_shear_distributions_tutorial.ipynb}}.
    \item The main \textsc{slsim} codebase, particularly:
    \begin{itemize}
        \item \textsc{main/slsim/Halos}.
        \item \textsc{main/slsim/LOS}.
    \end{itemize}
\end{itemize}

The external convergence and shear distribution datasets used in this work can be accessed at the public repository\footnote{\url{https://github.com/LSST-strong-lensing/data_public}}.

Researchers are encouraged to explore these resources to reproduce our results or adapt them for related studies.


\bibliographystyle{plainnat}
\bibliography{example}



\end{document}